\documentclass[sn-chicago, 12pt]{sn-jnl}

\usepackage{graphicx}%
\usepackage{multirow}%
\usepackage{amsmath,amssymb,amsfonts}%
\usepackage{amsthm}%
\usepackage{mathrsfs}%
\usepackage[title]{appendix}%
\usepackage{xcolor}%
\usepackage{textcomp}%
\usepackage{manyfoot}%
\usepackage{booktabs}%
\usepackage{algorithm}%
\usepackage{algorithmicx}%
\usepackage{algpseudocode}%
\usepackage{listings}%
\usepackage{dsfont}

\newcommand{\Y}{\mathbf{Y}}
\newcommand{\bu}{\mathbf{u}}

\newcommand{\Z}{\mathrm{Z}}

\newcommand{\bSigma}{\boldsymbol{\Sigma}}
\newcommand{\btheta}{\boldsymbol{\vartheta}}
\newcommand{\bTheta}{\boldsymbol{\Theta}}
\newcommand{\itert}{^{(t)}}
\newcommand{\itertm}{^{(t-1)}}

\newcommand{\pell}{^{(\ell)}}
\newcommand{\pk}{^{(k)}}
\makeatother

\theoremstyle{thmstyleone}%
\theoremstyle{thmstyletwo}%

\theoremstyle{thmstylethree}%

\begin{document}

\title[ALINT]{Spatially Regularized Gaussian Mixtures for Clustering Spatial Transcriptomic Data}


\author*[1]{\fnm{Andrea} \sur{Sottosanti}}\email{andrea.sottosanti@unipd.it}
\author[1]{\fnm{Davide} \sur{Risso}}\email{davide.risso@unipd.it}
\author[1]{\fnm{Francesco} \sur{Denti}}\email{francesco.denti@unipd.it}


\affil[1]{\orgdiv{Department of Statistical Sciences}, \orgname{University of Padova}, \orgaddress{\street{via Cesare Battisti 241}, \city{Padova}, \postcode{35121}, \country{Italy}}}


\abstract{
Spatial transcriptomics measures the expression of thousands of genes in a tissue sample while preserving its spatial structure. This class of technologies has enabled the investigation of the spatial variation of gene expressions and their impact on specific biological processes. Identifying genes with similar expression profiles is of utmost importance, thus motivating the development of flexible methods leveraging spatial data structure to cluster genes. Here, we propose a modeling framework for clustering observations measured over numerous spatial locations via Gaussian processes. Rather than specifying their covariance kernels as a function of the spatial structure, we use it to inform a generalized Cholesky decomposition of their precision matrices. This approach prevents issues with kernel misspecification and facilitates the estimation of a non-stationarity spatial covariance structure. Applied to spatial transcriptomic data, our model identifies gene clusters with distinctive spatial correlation patterns across tissue areas comprising different cell types, like tumoral and stromal areas.}


\maketitle

\clearpage

\section{Introduction}
\subsection{Spatial transcriptomic gene expression data analysis}
\label{subsec:data_structure}

In recent years, significant technological advances have led to the development of more efficient platforms for examining gene expression, including single-cell RNA sequencing (scRNA-seq) and, more recently, spatial transcriptomics. Spatial transcriptomics refers to methods that measure the expression of thousands of genes in a tissue sample while preserving its structural information. Unlike scRNA-seq, spatial transcriptomics enables the study of the spatial distribution of gene expressions across a tissue.

Spot-based spatial transcriptomic technologies \citep{Righelli_etal.2021} capture RNA molecules from small spots arranged on a grid over a tissue slide. The number of cells collected in each spot depends on the specific instrument used and its resolution. For instance, the \textit{Visium}\footnote{https://www.10xgenomics.com/platforms/visium} technology collects an average of 1 to 10 cells per spot, whereas \textit{Visium HD}\footnote{https://www.10xgenomics.com/products/visium-hd-spatial-gene-expression} offers subcellular resolution, allowing detection of gene expression variation within individual cells.

The advent of these technologies has led to the development of statistical methods to analyze gene expressions by leveraging the spatial information provided by these new tools. A key focus in the literature has been the detection and study of genes that exhibit specific patterns of spatial variation, commonly referred to as spatially variable genes. This problem has been addressed using testing procedures that analyze each gene individually \citep{Svensson_etal:2018, Edsgard_etal.2018, Sun_etal:2020, Wang_etal.2023, Weber_etal.2023}. 
The interest extends beyond simply identifying spatially expressed genes. For instance, \cite{Svensson_etal:2018} and \cite{Sun_etal:2020} employed different clustering methods on subsets of genes identified as spatially variable, aiming to identify groups of genes with similar spatial variation profiles. Identifying such clusters is fundamental to many biological analyses, including the discovery of gene expression signatures, i.e., groups of genes sharing specific expression patterns that can provide insights into underlying biological processes \citep{Itadani_etal.2008}.

\begin{figure}[t]
    \centering
    \includegraphics[width=\linewidth]{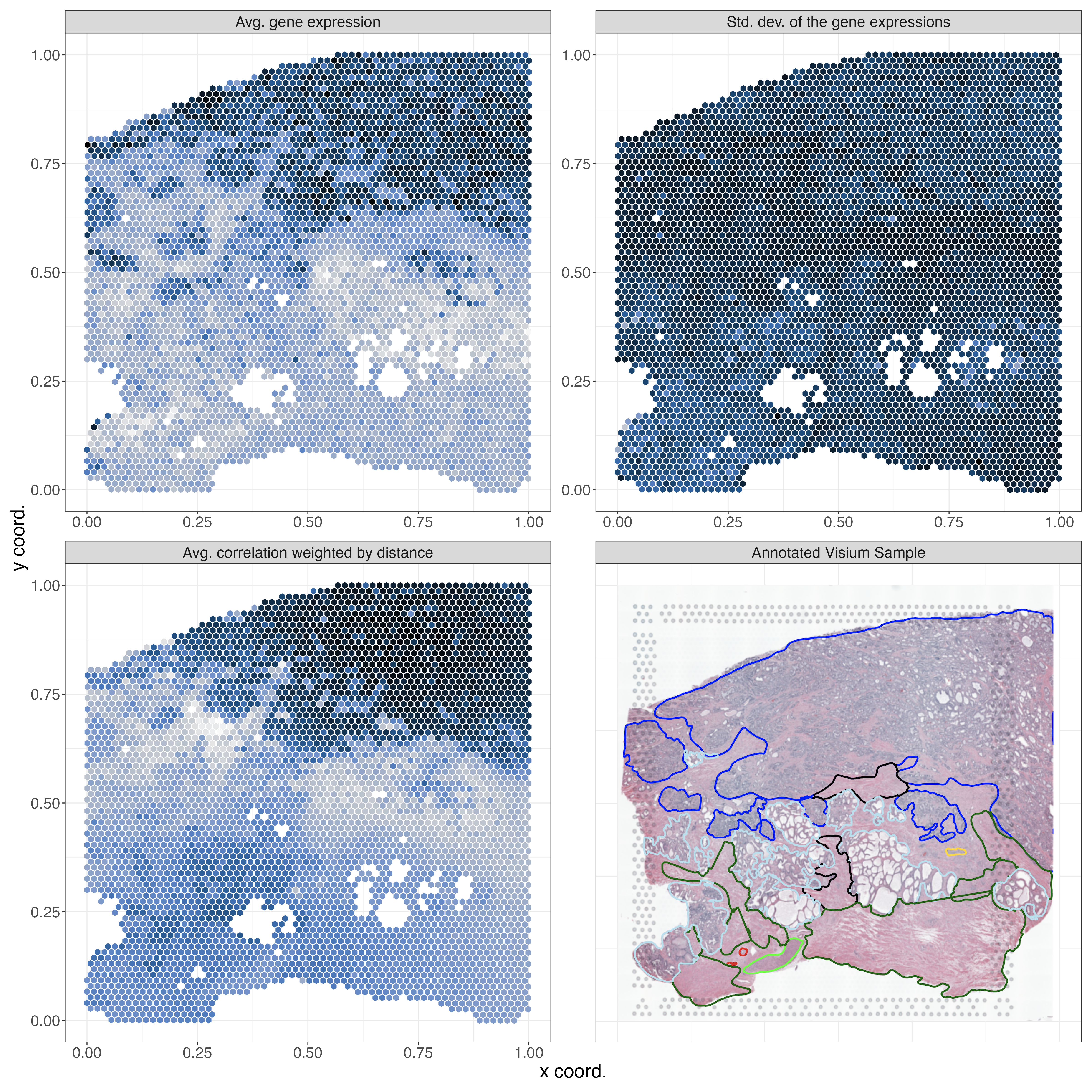}
    \caption{Example of a prostate cancer tissue sample analyzed with Visium. The first three panels illustrate the multifaceted nature of the gene expression data: the spot-wise average gene expressions (top-left), their spot-wise standard deviation (top-right), and the average correlation between each spot and its neighbors, reweighted for spatial (Euclidean) distance (bottom-left). All values are normalized. The bottom-right panel shows the openly available annotation of the Visium image. The main areas of interest are the invasive carcinoma (in blue), the fibro-muscular tissue (in dark green), and the fibrous tissue (in black).
}
    \label{fig:EDA}
\end{figure}

This work addresses the need for versatile and computationally efficient methods to cluster gene expressions based on their spatial variation profiles. To demonstrate our approach, we employ a human prostate tissue sample processed with Visium technology as a case study. The dataset, which is publicly available online\footnote{https://www.10xgenomics.com/datasets/human-prostate-cancer-adenocarcinoma-with-invasive-carcinoma-ffpe-1-standard-1-3-0}, contains expression data for over 30,000 genes measured across more than 4,300 spatially defined locations (called spots). This tissue section is derived from a human stage III prostate tumor and has been annotated with invasive adenocarcinoma. By identifying spatially coherent clusters, we aim to uncover biologically meaningful patterns and gain insight into the biological processes that drive prostate cancer progression and dynamics. For this analysis, we selected the top 500 most highly variable genes using the pre-processing method of \cite{Townes_etal.2019}. Further pre-processing details are provided in Section~\ref{sec:application}. Figure~\ref{fig:EDA} presents three heatmaps that illustrate distinct aspects of gene expression across the tissue spots.  The top-left panel illustrates the average gene expression, the top-right panel displays the standard deviation, and the bottom-left panel shows the average correlation between each spot and its neighbors, reweighted for spatial (Euclidean) distance. All values are normalized, with color gradients ranging from dark blue (representing the minimum) to white (representing the maximum), providing a first visual indication of the complexity of gene expression patterns across the tissue.
Lastly, in the bottom-right panel, we report the annotated slice showing the cell types of the considered tissue. These annotations are publicly available online from the same source of the data.

\subsection{Spatial covariance estimation in clustering}

In this article, we address the challenge of modeling and clustering gene expression profiles in tissue samples processed using spot-based spatial transcriptomic methods, with measurements obtained at spatially located sites.
    

Model-based clustering methods play a pivotal role in the literature, as they rely on probabilistic models to allocate observations into groups, providing not only a partition of the dataset but also quantifying the uncertainty associated with the cluster membership of each observation. Mixture models are particularly appealing in this framework, as they represent a generative mechanism that accounts for missing information -- specifically, the clustering labels. These labels are treated as unknown quantities and are inferred during the analysis. For a detailed review of model-based clustering, see, for example, \cite{Bouveyron_etal.2019}.


When observations are recorded at spatially located points, a clustering model must 
account for the dependence of measurements across these spatial points. A covariance matrix typically captures this dependence, where each entry quantifies the association between two spatial locations. Under stationarity assumptions, this spatial dependence can be modeled using various kernel functions, as thoroughly described, for example, by \cite{Rasmussen_Williams:2006}. When the data are assumed to follow a Gaussian distribution, a well-established solution is given by Gaussian processes \citep{Rasmussen_Williams:2006,Banerjee_etal.2015}. This flexible modeling framework represents any realization of a phenomenon over a finite set of spatial points as a multivariate normal distribution, with the covariance between spatial points defined by a selected kernel function. Gaussian processes provide a robust approach to capture spatial relationships and have been used extensively in many fields of applications, such as climate model emulation \citep{Castruccio_Stein.2013}, functional magnetic resonance imaging (fMRI, \citealt{Caponera_etal.2017}), and genetic association mapping \citep{Kumasaka.2024}.

A natural solution for clustering observations measured in space combines mixture models with Gaussian processes. For instance, one can construct a mixture model where each component is a multivariate normal distribution, with covariance matrices defined by kernel functions. The same kernel function can be used for all components or allowed to vary across them. This approach has recently been applied to clustering tasks, including functional data \citep{Paton_Mcnicholas.2020}, spatio-temporal data \citep{Vanhatalo_etal.2021}, and spatial data in a co-clustering framework \citep{sottosanti_risso.2023, Sottosanti_etal.2025}. In addition, this approach was used by \cite{Svensson_etal:2018} to cluster spatially expressed genes. However, a notable drawback of this methodology lies in its computational cost. Computing the covariance function requires $\mathcal{O}(n^3)$ operations and $\mathcal{O}(n^2)$ memory, where $n$ is the number of spatial locations. These costs are further multiplied by the number of mixture components. In spatial transcriptomics, where technological advances enable the analysis of increasingly larger tissue samples, such computational demands become prohibitive. To address this limitation, \cite{datta_etal.2016} proposed an innovative and efficient approach, known as nearest neighbor Gaussian processes (NNGP), which drastically reduces the computational costs by simplifying the dependence structure across spatial locations. In particular, they consider only a subset of neighbors for each point, allowing to resort sparse matrices. The method has gained significant attention in recent years (see, for example, \citealt{guinness.2018}, \citealt{Heaton_etal.2019}, \citealt{Katzfuss_Guinness.2021}). However, despite the promising computational and modeling advantages of NNGPs, their application within clustering frameworks is still in the early stages of development and, to the best of our
knowledge, has been under-explored \citep{sottosanti_etal.2023, chakraborty_Chakraborty.2024}.


An alternative approach involves constructing a mixture model in which the covariance matrices of each component are fully estimated from the data. Various forms of regularization can be applied to simplify the model and reduce computational costs. For example, the parsimonious Gaussian mixture modeling framework \citep{celeux_govaert.1995, Scrucca_etal.2016} imposes constraints on the eigen-decomposition of the covariance matrices, allowing for models with different levels of complexity. These constraints may require components to share the same shape, orientation, or volume. Additionally, covariance matrix estimation can be regularized through penalization techniques. For instance, methods proposed by \cite{huang_etal.2006} and \cite{Friedman_etal.2008} estimate covariance matrices in a nonparametric manner by imposing sparsity on their inverses, also known as precision matrices. 
While these approaches are flexible and computationally efficient, they are effectively disconnected from spatial modeling frameworks, as they do not incorporate spatial information to capture the dependence among measurement points.


In this work, we propose a novel modeling solution that bridges the gap between the fully parametric spatial approach of classical Gaussian processes and the fully data-driven methods described above. Our approach builds upon recent advancements in spatial modeling presented in \cite{kidd_katzfuss.2022}. The authors introduced a method that avoids imposing rigid structures on the covariance matrix by estimating the modified Cholesky decomposition of its inverse using Bayesian inference. Spatial information is incorporated through the prior distributions, which regularize the estimates while enabling flexible reconstruction of dependence across data points. This flexibility allows the approach to handle complex scenarios, such as non-stationarity in spatial processes and thus does not require the specification of a rigid kernel function. We integrate this method into the Gaussian mixture modeling framework, naming our approach {spatially regularized Gaussian mixture model} (SR-GMM). 


The rest of the manuscript is structured as follows. Section \ref{sec:method} revisits the covariance modeling approach of \cite{kidd_katzfuss.2022} and discusses its limitations in the context of spatial transcriptomic data due to its need for several replicates. This sets the stage for introducing the concept of clustering genes with similar spatial expression profiles and characteristics, which leads to the development of our SR-GMM framework. Section \ref{sec:inference} presents an algorithm that efficiently provides parameter estimates. Section \ref{sec:simulations} proposes a series of simulation experiments to test the ability of our model to recover the clustering structure of the data under various experimental conditions, including cases where the SR-GMM is misspecified due to additional sources of variability in the data. Section \ref{sec:application} demonstrates the application of our SR-GMM to the prostate cancer tissue sample described in Section \ref{subsec:data_structure}, showcasing the model's ability to identify clusters of genes with specific spatial dependence structures. This section highlights the advantages of using a flexible, non-parametric covariance model, especially in identifying and visualizing regions of both strong and weak spatial dependence within the tissue. An interpretation of the discovered clusters is also provided through gene set enrichment analysis \citep{khatri2012ten}. Finally, Section \ref{sec:discussion} offers future perspectives and research directions for further extending the model.
    

\section{Methodology}
\label{sec:method}

\subsection{Notation}
First, each spatial location $i=1,\ldots,n$ (also known as spot) is characterized by its spatial coordinates $\mathbf{s}_i$, which are all collected in the matrix $\mathbf{S}=\{\mathbf{s}_i\}_{i = 1}^{n}$. Thus, $\mathbf{S}$ has dimension $n\times 2$. In our context, the data collected from a spatial experiment conducted on a biological tissue are organized in a matrix $\Y$, of dimension $N\times n$. We denote the single entry of this matrix with $y_{\ell,i}$, which indicates a measure of the gene expression of the $\ell$-th gene, with $\ell=1,\dots,N$, recorded in the spatial location $i$, with $i=1,\ldots,n$. We will refer to $\mathbf{y}_{\ell,.}$ as the vector collecting of expression of the $\ell$-th gene across the $n$ spatial spots that collect the tissue cells. 

In general, when dealing with any generic matrix $\mathbf{A}$ of size $p \times q$,  $\mathbf{A}_{i,.}$ will denote its $i$-th row vector of length $q$, and $\mathbf{A}_{.,j}$ will denote its $j$-th column vector of length $p$. Let $\mathcal{I}$ and $\mathcal{J}$ represent two subsets of row and column indices, where $\mathcal{I}\subset \{1,\ldots,p\}$ and $\mathcal{J}\subset \{1,\ldots,q\}$. Then, $\mathbf{A}_{\mathcal{I},\mathcal{J}}$ represents the submatrix obtained considering the corresponding rows and columns. An example of subset we will use later is given by $a:b$, a compact notation for $\{a,a+1,\ldots,b-1,b\}$, for two integers $a<b$.

Since our goal is to partition genes into groups of instances that exhibit similar nonstationary spatial dependence, we will also need to refer to gene- and cluster-specific quantities. Therefore, let us represent with $\mathbf{b}^{(\ell)}$ and $\mathbf{A}^{(\ell)}$ a generic vector and matrix pertaining the $\ell$-th gene, respectively. Then, to access the $(p,q)$-th element of the matrix, we will use the notation $\mathrm{A}\pell_{p,q}$. Similarly, when referring to quantities specific of the $k$-th cluster, we will use $\mathbf{b}^{(k)}$, $\mathbf{A}^{(k)}$, and $\mathrm{A}^{(k)}_{p,q}$.

In the following, we first outline the method proposed by \citet{kidd_katzfuss.2022}, using it to model the expression of every gene while introducing the concept of spatial regularization in estimating the covariance. To enhance estimation by leveraging information across genes and to induce data clustering, we then extend the model by introducing a novel class of mixtures, which we refer to as spatially regularized Gaussian mixture models (SR-GMM).

\subsection{Non-stationary covariance model for spatial gene expressions}
\label{subsec:flexible_model}

First, we assume that the entries of $\Y$ have been pre-processed in a way that $y_{\ell,i}\in\mathbb{R}$ and the distribution of the values in $\mathbf{y}_{\ell,.}$ is nearly symmetric. This can be achieved, for example, using the method described by \cite{Townes_etal.2019}. Further details on the application of this method to spatial transcriptomic data are provided in Section \ref{sec:application}, in the context of a case study analysis. According to \cite{Weber_etal.2023} and to the related literature, we can express, for $\ell=1,\ldots,N$, the following  multivariate normal model:
\begin{equation}
    \label{formula:multivariate_gaussian_single_gene}
\mathbf{y}_{\ell,.}\sim\mathcal{N}(\boldsymbol{\mu}\pell,\bSigma\pell),
\end{equation}
where $\boldsymbol{\mu}\pell$ can be either constant or can include eventual exogenous variables. 
In Equation~\eqref{formula:multivariate_gaussian_single_gene}, $\boldsymbol{\mu}\pell$ and $\bSigma\pell$ denote a mean vector and a covariance matrix specific to the $\ell$-th observation.
Without loss of generality, we consider the case $\boldsymbol{\mu}\pell = \mu_\ell\mathbf{1}_n$, with $\mu_\ell \in \mathbb{R}$. Differences in gene mean expressions may result from biological factors (e.g., tissue type) or technical artifacts (e.g., sample preparation or batch effects). A common approach in the literature \citep{Svensson_etal:2018, Sun_etal:2020, Weber_etal.2023} consists in defining the covariance matrix $\bSigma\pell$ as a function of distance between spatial data points through a parametric kernel function, therefore setting, for two spatial locations $i$ and $i'$,  $\Sigma\pell_{i,i'}=\kappa(||\mathbf{s}_{i}-\mathbf{s}_{i'}||;\boldsymbol{\Omega}\pell)$, where $\kappa(\cdot;\cdot)$ is a parametric spatial covariance function parametrized by $\boldsymbol{\Omega}\pell$ \citep{Rasmussen_Williams:2006}. 
Despite its large utility, this approach is heavily limited by its lack of flexibility, especially when the data show evidence of non-stationarity. These and other limitations have led to the development of nonparametric methods (see, for example, \citealt{Wang_etal.2023}). To achieve a larger degree of flexibility, we consider the generalized Cholesky decomposition of $\bSigma\pell$, that is 
\begin{equation}
    \label{formula:Cholesky_factorization_single_gene}
    (\mathbf{U}\pell)^\top\bSigma\pell\mathbf{U}\pell=\mathbf{D}\pell,
\end{equation}
where $\mathbf{D}\pell = \mathrm{diag}(d\pell_{1},\ldots,d\pell_{n})$ is a positive diagonal matrix and $\mathbf{U}\pell$ is an upper triangular matrix having ones on the main diagonal. Such decomposition allows to write  the precision matrix $(\bSigma^{(\ell)})^{-1}$ as
$$
(\bSigma^{(\ell)})^{-1} = \mathbf{U}^{(\ell)}(\mathbf{D}^{(\ell)})^{-1}(\mathbf{U}^{(\ell)})^\top.
$$
If $\bSigma\pell$ is specified through a parametric kernel as illustrated above, then the quantities $\mathbf{U}\pell$ and $\mathbf{D}\pell$ are deterministically derived. As shown for example by \cite{huang_etal.2006}, the generalized Cholesky decomposition of $\bSigma\pell$ allows to rewrite the joint distribution of $\mathbf{y}_{\ell,.}$ as a sequence of univariate regression models, that is
\begin{equation}
\label{formula:sequence_of_regressions}
y_{\ell,i}|\tilde{\mathbf{y}}_{\ell,1:(i-1)}\sim\mathcal{N}(\mu_\ell-\tilde{\mathbf{y}}_{\ell,1:(i-1)}\mathbf{U}\pell_{1:(i-1),i},d\pell_i),
\end{equation}
for $i>1$, where $\tilde{\mathbf{y}}_{\ell,.} = \mathbf{y}_{\ell,.}-\mu_\ell\mathbf{1}_{n}$; in addition, $y_{\ell,1}\sim\mathcal{N}(\mu_\ell,d\pell_1)$. 

In Formula~\eqref{formula:sequence_of_regressions}, the number of regression coefficients increases with $i$, leading to a substantial growth of the computational cost in terms of the required number of operations and memory. The regressions can be approximated assuming that $y_{\ell,i}$ depends on at most $m_\ell$ neighbors, where $m_\ell \ll n - 1$. This construction, known as the Vecchia approximation \citep{vecchia.1988}, has been extensively explored during the last decade to induce sparsity in spatial data analysis in the presence of a large number of spatial points \citep{datta_etal.2016, Saha_Datta.2018}.
Being $g_{m_\ell}(i) \subset \{1,\dots,i-1\}$ a set of size $m_{\ell,i} = \min(m_\ell, i-1)$,  we rewrite the regressions in \eqref{formula:sequence_of_regressions} for which $i>1$ as
\begin{equation}
    \label{formula:regression2}
    y_{\ell,i}|\tilde{\mathbf{y}}_{\ell,g_{m_\ell}(i)}\sim\mathcal{N}(\mu_\ell-\tilde{\mathbf{y}}_{\ell,g_{m_\ell}(i)}\bu\pell_i,d\pell_i),
\end{equation}
where, for simplicity, we write $\bu\pell_{i} = \mathbf{U}\pell_{g_{m_\ell}(i),i}$. In a fully nonparametric approach, the set $g_{m_\ell}(i)$ can be determined directly from the data, for instance, selecting $m_{\ell,i}$ coefficients among the $i-1$ in $\mathbf{U}\pell_{1:(i-1),i}$ through a LASSO-type penalization \citep{huang_etal.2006}. Alternatively, $g_{m_\ell}(i)$ may represent a set of spatial neighboring points that forerun $i$, determined by the spatial distribution of the spots. It becomes clear that, in this framework, how spatial points are ordered becomes crucial, as spatial coordinates lack an inherent ordering. While Formula~\eqref{formula:sequence_of_regressions} remains unaffected by this aspect -- any arrangement of points results in the same joint distribution of $ \mathbf{y}_{\ell,.} $ -- the formulation in \eqref{formula:regression2} introduces an approximation that makes the ordering significant. Hereon, we assume that the experimental spots, represented as the rows of $ \mathbf{S} $, are ordered using the maximin method, which has been shown empirically to outperform other sorting techniques for approximating a joint density function \citep{guinness.2018, Schafer_etal.2021}. 
Recently, \cite{kidd_katzfuss.2022} proposed to infer $(\bu\pell_i,d\pell_i)$ through a hierarchical structure, thus assuming a normal-Inverse Gamma specification:
\begin{equation}
\label{formula:normal_inverse_gamma}
    \bu\pell_i|d\pell_i\sim\mathcal{N}(\mathbf{0},d\pell_i\mathbf{V}\pell_i),\hspace{1cm}d\pell_i\sim\mathcal{IG}(\alpha\pell_i,\beta\pell_i).
\end{equation}
In Equation~\eqref{formula:normal_inverse_gamma}, $\mathbf{V}\pell_i = \mathrm{diag}(\mathrm{V}\pell_{i,1},\dots,\mathrm{V}\pell_{i,m_{\ell,i}})$ is a diagonal covariance matrix of dimension $m_{\ell,i}$, specific of the $\ell$-th observation and the $i$-th spot, and $\mathcal{IG}(a,b)$ denotes the inverse-gamma distribution with shape $a$ and scale $b$. 
To embed the spatial information into the model, the hyperparameters are reparametrized as
$$
\mathrm{V}\pell_{i,i'} = \dfrac{\exp(-\vartheta\pell_{3}i')}{\vartheta\pell_{1}f(i, \vartheta\pell_{2})},\hspace{.7cm}\alpha\pell_i  = 6,\hspace{.7cm}
\beta\pell_i = 5\vartheta\pell_{1}f(i, \vartheta\pell_{2}),
$$
where $i'=1,\dots,m_{\ell,i}$ and $f(i, \vartheta_{2}) = 1-\exp(-\vartheta_{2}/\sqrt{i})$. The positive parameters $\boldsymbol{\vartheta}\pell = (\vartheta\pell_{1}, \vartheta\pell_{2}, \vartheta\pell_{3})$  are thought to represent the marginal variability, the range, and the smoothness of the spatial processes. 

Finally, to select the number of neighbors $m_{\ell}$, \cite{kidd_katzfuss.2022} proposed to set $m_\ell$ equal to the largest $i'\in \{1,\dots,n\}$ such that the expected variation a priori of $\mathbf{u}\pell_i$, $\exp(-\vartheta\pell_{3}i')$, remains above a small threshold, e.g., 0.001, which represents the minimum amount of variation that can be detected in the data.

\subsection{Clustering genes via spatially regularized Gaussian mixtures}
\label{subsec:clustering_model}

The model introduced in Section \ref{subsec:flexible_model} recovers the spatial covariance structure of each gene, therefore aligning with several established methods for spatial transcriptomic data analysis (e.g., \citealt{Svensson_etal:2018}, \citealt{Sun_etal:2020}, \citealt{Weber_etal.2023}). Nonetheless, because our model is based on the method of \cite{kidd_katzfuss.2022}, it encounters limitations due to its reliance on repeated spatial observations to recover $\mathbf{U}$ and $\mathbf{D}$. To overcome this limitation, we propose a modeling framework that leverages shared information across multiple genes by grouping those with similar spatial expressions, enabling clustering based on different spatial correlation profiles, which are nonparametrically estimated.  In addition, the clustering accounts for differences in overall mean gene expression levels across clusters.

We tackle this clustering problem by introducing, for every gene $\ell$, a sequence of binary random variables $\Z_{\ell, k} = \mathds{1}(\ell \in \mathcal{C}_k)$ to denote whether $\ell$ belongs or not to the $k$-th cluster, denoted with $\mathcal{C}_k$, for $k = 1,\dots,K$. The size of $\mathcal{C}_k$ is $N_k = \sum_{\ell=1}^N\Z_{\ell,k}$. 
Model \eqref{formula:multivariate_gaussian_single_gene} is thus extended to
\begin{equation}
    \label{formula:conditional_model}
    \mathbf{y}_{\ell,.}|\Z_{\ell,k}= 1 \sim\mathcal{N}(\mu_k\mathbf{1}_n,\bSigma\pk),\hspace{.3cm}(\bSigma\pk)^{-1} = \mathbf{U}\pk(\mathbf{D}\pk)^{-1}(\mathbf{U}\pk)^\top,
\end{equation}
where $\mathbf{Z}_{\ell,.}|\boldsymbol{\pi}$ is distributed according to a multinomial random variable with probability vector $\boldsymbol{\pi}$, i.e., $\mathbb{P}\left({\Z}_{\ell,k}=1\right)=\pi_k$, $\forall k$. Recall that the superscript $(k)$ indicates a quantity specific of the $k$-th cluster, therefore shared across all the observations in $\mathcal{C}_k$. Similarly to Equation~\eqref{formula:normal_inverse_gamma_cluster}, the distribution of the quantities $(\mathbf{U}\pk, \mathbf{D}\pk)_{k = 1,\dots,K}$ is defined via
\begin{equation}
\label{formula:normal_inverse_gamma_cluster}
    \bu\pk_{i}|d\pk_{i}\sim\mathcal{N}(\mathbf{0},d\pk_{i}\mathbf{V}\pk_{i}),\hspace{1cm}d\pk_{i}\sim\mathcal{IG}(\alpha\pk_{i},\beta\pk_{i}).
\end{equation}
Recall that the quantities $(\alpha\pk_{i},\beta\pk_{i},\mathbf{V}\pk_{i})$ are deterministic transformations of the elements of $\btheta\pk$. 
We denote with $\bTheta_k = (\mu_k, \btheta\pk, \pi_k)$ the set of model parameters referred to the $k$-th cluster.  The model~\eqref{formula:conditional_model} represents a mixture of multivariate normal densities with covariance matrices that are not constrained by a rigid parametric kernel. Instead, they are estimated from the data through a hierarchical specification incorporating the data’s ordering and neighboring structure. Given these characteristics, we refer to \eqref{formula:conditional_model}-\eqref{formula:normal_inverse_gamma_cluster} as spatially regularized Gaussian mixture model (SR-GMM).

\begin{figure}
    \centering
    \includegraphics[width=0.6\linewidth]{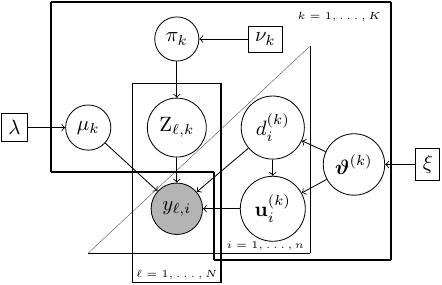}
    \caption{Directed acyclic graph (DAG) of the proposed spatially regularized Gaussian mixture model. The grey circle denotes the data, the white circles denote the model parameters, and the white rectangles denote the hyperparameters.}
    \label{fig:DAG}
\end{figure}

In a fully Bayesian framework, the model specification is completed by defining the a priori distributions for $\bTheta_k$, e.g.,
$$
\mu_k\sim\mathcal{N}(0,  \lambda),\hspace{1cm}\btheta\pk\sim\log\mathcal{N}(\mathbf{0},\xi\;\mathbf{I}_3),\hspace{1cm}\boldsymbol{\pi}\sim \text{Dirichlet}_K(\nu_1,\dots,\nu_K),
$$
where $\log \mathcal{N}(\mathbf{a},\mathbf{B})$ denotes the multivariate log-normal distribution with location vector $\mathbf{a}$ and scale matrix $\mathbf{B}$. The first two prior distributions are meant to impose an $L_2$ regularization on $\mu_k$ and $\log(\btheta\pk)$ via $\lambda$ and $\xi$: the first helps ensure that clustering is driven not only by mean levels but also by the spatial covariance structure, while the second prevents excessively small or large values of the $\boldsymbol{\vartheta}$ parameters, which could lead to unstable results. Lastly, the $K$-dimensional Dirichlet distribution is selected for its conjugacy with the categorical model. In alternative to the  Bayesian approach, one can perform a frequentist estimation of $\bTheta_k$, as also considered by \cite{kidd_katzfuss.2022}. Figure~\ref{fig:DAG} represents the dependence structure of the model parameters through a directed acyclic graph (DAG), assuming the Bayesian model specification.

To conclude this section, we remark that one might consider an alternative formulation of the SR-GMM, compromising between model~\eqref{formula:multivariate_gaussian_single_gene}  and  \eqref{formula:conditional_model}. Indeed, one can specify a model where a different covariance matrix $ \bSigma\pell $ is assigned to each gene while using cluster-specific parameters $ \btheta\pk $ instead of gene-specific parameters $ \btheta^{(\ell)} $:
\begin{equation*}
\begin{gathered}
\mathbf{y}_{\ell,.}|\Z_{\ell,k}= 1,\mu_k,\btheta\pk \sim\mathcal{N}\left(\mu_k\mathbf{1}_n,\bSigma\pell\right),\hspace{1cm}\bSigma\pell = (\mathbf{U}^\top_\ell)^{-1}\mathbf{D}\pell\mathbf{U}^{-1}_\ell,\\
\bu\pell_i|d\pell_i,\Z_{\ell,k}=1\sim\mathcal{N}(\mathbf{0},d\pell_i\mathbf{V}\pk_{i}),\hspace{1cm}d\pell_{i}|\Z_{\ell,k}=1\sim\mathcal{IG}(\alpha\pk_{i},\beta\pk_{i}).
\end{gathered}
\end{equation*}

Under this setup, observations within a cluster do not share the same covariance matrix but, instead, the distribution that generates the covariance matrix, for each gene. 
This formulation is particularly attractive due to its flexibility. It leads to several advantages in the estimation phase, as it allows for parallel cluster observations (more details on the estimation phase are provided in Section~\ref{sec:inference}). 

However, our model \eqref{formula:conditional_model}-\eqref{formula:normal_inverse_gamma_cluster} recovers $\bSigma$ in a non-parametric fashion, quantifying the dependence across spatial points directly from the data with a prior distribution that incorporates the spatial proximity. Therefore, multiple genes must inform not only the estimation of the prior parameters $\btheta$ but also of the covariance $\bSigma$. If $\bSigma$ were specified with a parametric kernel, then clustering based on the kernel’s parameters might be appropriate, as the kernel imposes a rigid structure to the covariance across spatial locations. We conclude that the formulation of the SR-GMM presented in Formulas  \eqref{formula:conditional_model}-\eqref{formula:normal_inverse_gamma_cluster} appears to be more adequate to the task of clustering multivariate observations measured in space without directly specifying their spatial covariance structure.

\section{Efficient posterior inference via stochastic expectation maximization}
\label{sec:inference}
The posterior estimation of the clustering model presented in Section~\ref{subsec:clustering_model} can quickly become computationally expensive if tackled via traditional MCMC simulation. Given the dimensionality of the problem and the need for timely results, we seek to obtain a point estimate of the model parameters $\bTheta = (\bTheta_k)_{k = 1,\dots,K}$ and the clustering labels $\mathbf{Z}$. This approach aligns with both the frequentist inferential framework, where the parameters are directly estimated from the data, and with the Bayesian framework, where parametric priors regularize the parameter estimates. 
In the former case, the estimation procedure targets the values that maximize the log-likelihood function. In contrast, in the latter case, the objective function to be maximized is the posterior density of $\bTheta$. In this section, we outline the estimation procedure considering a fully Bayesian approach. The procedure can be adapted to the empirical Bayes framework with straightforward modifications.

To obtain our estimation algorithm, we start by writing the complete likelihood of the model as
$$ \mathcal{L}(\Y,\mathbf{U},\mathbf{D},\mathbf{Z}|\bTheta) = \prod_{k=1}^K \mathcal{L}(\Y,\mathbf{U}\pk,\mathbf{D}\pk,\mathbf{Z}_{.,k}|\bTheta_k),$$
where we denote with
$ \mathcal{L}(\Y,\mathbf{U}\pk,\mathbf{D}\pk,\mathbf{Z}_{.,k}|\bTheta_k)$ the contribution of the $k$-th cluster to the complete likelihood function. This quantity can be written as 
\begin{equation}
\begin{aligned}
\label{formula:likelihood_k_contribution}
\mathcal{L}(\Y,\mathbf{U}\pk,\mathbf{D}\pk,\mathbf{Z}_{.,k}|\bTheta_k)=&\prod_{\ell= 1}^N \left\{\mathcal{N}\left(\mathbf{y}_{\ell,.}|\mu_k\mathbf{1}_n,\bSigma\pk\right)\right\}^{\Z_{\ell,k}}\{\pi_k\}^{\Z_{\ell,k}} \times\\&\prod_{i=1}^n \mathcal{N}\left(\bu\pk_{i}|\mathbf{0},d\pk_{i}\mathbf{V}\pk_{i}\right)\mathcal{IG}(d\pk_{i}|\alpha\pk_{i},\beta\pk_{i}),
\end{aligned}
\end{equation}
where $(\mathbf{U}\pk)^\top\bSigma\pk \mathbf{U}\pk = \mathbf{D}_k$. We further notice that the first term of the right-hand-side of Formula~\eqref{formula:likelihood_k_contribution} can be rewritten as
\begin{align*}
\prod_{\ell= 1}^N \left\{\mathcal{N}\left(\mathbf{y}_{\ell,.}|\mu_k\mathbf{1}_n,\bSigma\pk\right)\right\}^{\Z_{\ell,k}} &=  \prod_{i=1}^n\left\{\prod_{\ell= 1}^N  \mathcal{N}\left(y_{\ell,i}|\mu_k-\tilde{\mathbf{y}}_{\ell,g_{m_k}(i)}\mathbf{u}\pk_{i},d\pk_{i}\right)^{\Z_{\ell,k}}\right\}\\
&=\prod_{i=1}^n \mathcal{N}\left(\mathbf{y}_{\mathcal{C}_k,i}|\mu_k\mathbf{1}_{N_k}+\mathbf{X}\pk_{i}\mathbf{u}\pk_{i},d\pk_{i}\mathbf{I}_{N_k}\right),
\end{align*}
where $\mathbf{y}_{\mathcal{C}_k,i}$ denotes the expression of the genes assigned the $k$-th cluster in the $i$-th spot, and $\mathbf{X}\pk_{i}=-\tilde{\Y}_{\mathcal{C}_k,g_{m_k}(i)}=-(\Y_{\mathcal{C}_k,g_{m_k}(i)}-\mu_k\mathbf{1}_{N_k\times m_{k,i}})$. Integrating out the random quantities $(\mathbf{U}\pk,\mathbf{D}\pk)$ from  \eqref{formula:likelihood_k_contribution}, we obtain
\begin{equation}
    \label{formula:integrated_likelihood_k}
    \mathcal{L}(\mathbf{Y},\mathbf{Z}_{.,k}|\bTheta_k) \propto \left(\prod_{\ell= 1}^N \pi_k^{\Z_{\ell,k}}\right)\prod_{i=1}^n 
    \left\{
    \dfrac{|\mathbf{G}\pk_{i}|^{1/2}}{|\mathbf{V}\pk_{i}|^{1/2}}
    \dfrac{(\beta\pk_{i})^{\alpha\pk_{i}}}{(\tilde{\beta}\pk_{i})^{\tilde{\alpha}\pk_{i}}}
    \dfrac{\Gamma(\tilde{\alpha}\pk_{i})}{\Gamma(\alpha\pk_{i})}
    \right\},
\end{equation}
where 
$\mathbf{G}\pk_{i} = \{(\mathbf{X}\pk_{i})^\top\mathbf{X}\pk_{i}+(\mathbf{V}\pk_{i})^{-1}\}^{-1}$, $\tilde{\alpha}\pk_{i} = \alpha\pk_{i}+N_k/2$, 
and $\tilde{\beta}\pk_{i} = \beta\pk_{i}+(\mathbf{y}_{\mathcal{C}_k,i}-\mu_k\mathbf{1}_{N_k})^\top\{\mathbf{I}_{N_k}+\mathbf{X}\pk_{i}\mathbf{V}\pk_{i}(\mathbf{X}\pk_{i})^\top\}(\mathbf{y}_{\mathcal{C}_k,i}-\mu_k\mathbf{1}_{N_k})/2$.
We therefore express the posterior distribution of the model parameters and the clustering labels as
$$
p(\bTheta, \mathbf{Z}|\mathbf{Y}) = \prod_{k = 1}^K\mathcal{L}(\mathbf{Y},\mathbf{Z}_{.,k}|\bTheta_k)p(\bTheta_k).
$$

Although Formula \eqref{formula:integrated_likelihood_k} avoids the estimation of $(\mathbf{U}\pk, \mathbf{D}\pk)$, and thus reduces the computational burden of the whole estimation procedure, it does not allow to make the conditional distribution of the clustering labels $\mathbf{Z}$ explicit. This aspect notably complicates the derivation of an expectation maximization (EM) algorithm. In this work, we consider a stochastic version of the algorithm, known as Stochastic EM \citep{Celeux_Diebolt.1985, Celeux_Govaert.1992}, that alternates a simulation phase, which draws a clustering configuration from the conditional distribution $p(\mathbf{Z}|\bTheta,\mathbf{Y})$, to a maximization step, that updates the estimates of $\bTheta$ based on the simulated clustering configuration. The procedure is repeated until convergence.

\subsection{Stochastic EM algorithm}
In detail, let $(\mathbf{Z}, \boldsymbol{\Theta})\itertm$ be the clustering configuration and the parameter values obtained at the iteration $t-1$. At iteration $t$, our estimation algorithm articulates in the following iterative steps.
\begin{itemize}
    \item \textbf{SE Step}: when $\ell = 1$, let $\tilde{\mathbf{Z}} = \mathbf{Z}^{(t-1)}$, while when $\ell \in \{2,\dots,n\}$, let $\tilde{\mathbf{Z}}$ be the matrix with $\tilde{\mathbf{Z}}_{1:(\ell-1),.} = {\mathbf{Z}}^{(t)}_{1:(\ell-1),.}$ and $\tilde{\mathbf{Z}}_{\ell:N,.} = {\mathbf{Z}}^{(t-1)}_{\ell:N,.}$. Then, for $\ell = 1,\dots,N$ and $k = 1,\dots,K$ compute
    $$
    \Pr(\Z_{\ell,k} = 1|\tilde{\mathbf{Z}}_{-\ell,.},\bTheta\itertm,\Y)=\tilde{\pi}_{\ell,k}\propto
    \prod_{j = 1}^K \mathcal{L}\left(\Y,\tilde{\mathbf{Z}}^{(\ell,k)}_{.,j}|\bTheta_j\itertm\right) p(\bTheta\itertm_j),
    $$
    where $\tilde{\mathbf{Z}}_{-\ell,.}$ denotes the matrix $\tilde{\mathbf{Z}}$ without the $\ell$-th row, and  $\tilde{\mathbf{Z}}^{(\ell,k)}$ denotes the matrix that equals to $\tilde{\mathbf{Z}}$, but with the $\ell$-th row vector having 1 in position $k$ and zero elsewhere. Notice that the quantity $\mathcal{L}\left(\Y,\tilde{\mathbf{Z}}^{(\ell,k)}_{.,j}|\bTheta_j\itertm\right) $ can be computed using Formula \eqref{formula:integrated_likelihood_k}. Then, $\mathbf{Z}\itert_{\ell,.}$ is updated by drawing a value from the categorical distribution with probability vector $(\tilde{{\pi}}_{\ell,1},\dots,\tilde{{\pi}}_{\ell,K})$.
    \item \textbf{M Step}: given the updated clustering labels $\mathbf{Z}\itert$, the model parameters of the $k$-th mixture component,  $\bTheta_k = (\mu_k, \btheta\pk,\pi_k)$, are updated with the values that maximize 
    $$
  \arg \max_{\bTheta_k}\mathcal{L}(\mathbf{Y},\mathbf{Z}\itert_{.,k}|\bTheta_k)p(\bTheta_k),
    $$
    therefore taking their maximum a posteriori (MAP) of $\bTheta_k$ based on the data assigned to the $k$-th cluster. While the optimization of  $\mu_k$ and $\btheta\pk$ cannot be achieved in closed form, and thus require a numerical procedure to maximize $p(\mu_k,\btheta\pk|\mathbf{Z}\itert_{.k},\Y)$, the mixture weights are updated using the modal value of the conditional posterior distribution of $p(\boldsymbol{\pi}|\mathbf{Z}\itert)$, which is still a Dirichlet:
    $$
    \pi\itert_k=\dfrac{N\itert_k+\nu_k-1}{N+\sum_{k=1}^K\nu_k-K}.
    $$
\end{itemize}

The presented algorithm has some similarities with the Gibbs sampling algorithm. Both methods include a simulating step from the full conditional posterior distribution of the cluster membership labels. However, the Gibbs sampler continues to simulate all the remaining parameters from their respective full conditional distributions. Our proposed SEM finds the parameter estimates that maximize the posterior density given a certain partition of the data. While the stochastic allocation step is not guaranteed to increase the value of the posterior density value at each iteration, it generates an irreducible Markov chain with a unique stationary distribution, which is expected to be concentrated around the maximum a posteriori of $\boldsymbol{\Theta}|\Y, \mathbf{Z}$ \citep{Keribin_etal.2015}. We implemented a convergence criterion to terminate the algorithm when the increase in the logarithm of the posterior density value remains below a specified threshold for a predefined number of consecutive iterations. The final estimate of $(\mathbf{Z}, \bTheta)$ corresponds to the values obtained at the iteration where the posterior density achieves its maximum value.

In principle, the SE Step could have been performed in other ways. For example, one might generate a candidate clustering configuration $\mathbf{Z}^*$ by modifying the configuration from the previous iteration through specific proposal moves (e.g., \citealt{Nobile_Fearnside.2007}) and then accept or reject it using the Metropolis-Hastings algorithm. However, this approach becomes increasingly complex and computationally expensive as the sample size grows. Another alternative is to perform the update of each $\mathbf{Z}_{\ell,.}$ in parallel, leveraging the fact that the model assumes genes are conditionally independent given $(\mathbf{U}, \mathbf{D})$. This strategy, however, would require explicit simulation of the random parameters $(\mathbf{U}, \mathbf{D})$. The resulting augmented procedure would involve first sampling from $\mathbf{U}, \mathbf{D} \mid \mathbf{Y}, \mathbf{Z}, \bTheta$ using the results of \cite{kidd_katzfuss.2022}, followed by drawing cluster labels from $\mathbf{Z} \mid \mathbf{U}, \mathbf{D}, \mathbf{Y}, \bTheta$. However,  simulating $(\mathbf{U}, \mathbf{D}, \mathbf{Z})$ would substantially increase the variability of the algorithm.
In contrast, our approach relies on the integrated likelihood in Formula \eqref{formula:integrated_likelihood_k},  which allows each $\mathbf{Z}_{\ell,.}$ to be updated via a Gibbs move without requiring the simulation of additional parameters. However, this formulation does not permit the decomposition of the likelihood into contributions from individual observations. As a result, the updates of $\mathbf{Z}_{\ell,.}$ are conditional on one another, leading to a collapsed Gibbs sampling step~\citep{Liu.1994}. We emphasize that all these sampling strategies are designed to target the conditional distribution of $\mathbf{Z}$.

\section{Simulation experiments}
\label{sec:simulations}

We design a series of simulation studies to assess the performance of our proposed methodology under various experimental conditions. Specifically, we aim to determine whether the SR-GMM can accurately identify the clustering structure of the data when different spatial kernels are used to generate different subpopulations. Furthermore, we investigate the impact of an additional source of variability unrelated to the spatial structure, referred to as nugget effects, on the clustering results. 

Each simulated dataset consists of two clusters: observations in the first cluster are generated from a zero-mean multivariate normal distribution with a variance-covariance matrix defined by an exponential kernel. In contrast, data in the second cluster are generated using a zero-mean multivariate normal with a Gaussian kernel covariance function. Therefore, for any distance $d$ between two points in space, the two covariance functions take the following form:
\begin{equation}
    \label{formula:simulation_kernels}
\psi_1(d;\sigma^2_1,\phi_1) = \sigma^2_1 \exp\left(-\dfrac{d}{\phi_1}\right),\hspace{.9cm}
\psi_2(d;\sigma^2_2,\phi_2) = \sigma^2_2 \exp\left(-\dfrac{d^2}{\phi^2_2}\right).
\end{equation}
The baseline configuration we investigate assumes equal values for the variance parameters, $\sigma^2_1 = \sigma^2_2 = 2$, and the scale parameters, $\phi_1 = \phi_2 = 1$. This setup defines a framework where the spatial expressions of the two clusters differ only in their parametric structure while sharing the same variance and scale. Note that the differences between the two clusters are purely given by the correlation structure, as they are not amplified by mean shifts. This framework serves as an effective testing ground for our methodology.
\begin{figure}
    \centering
    \includegraphics[width=\textwidth]{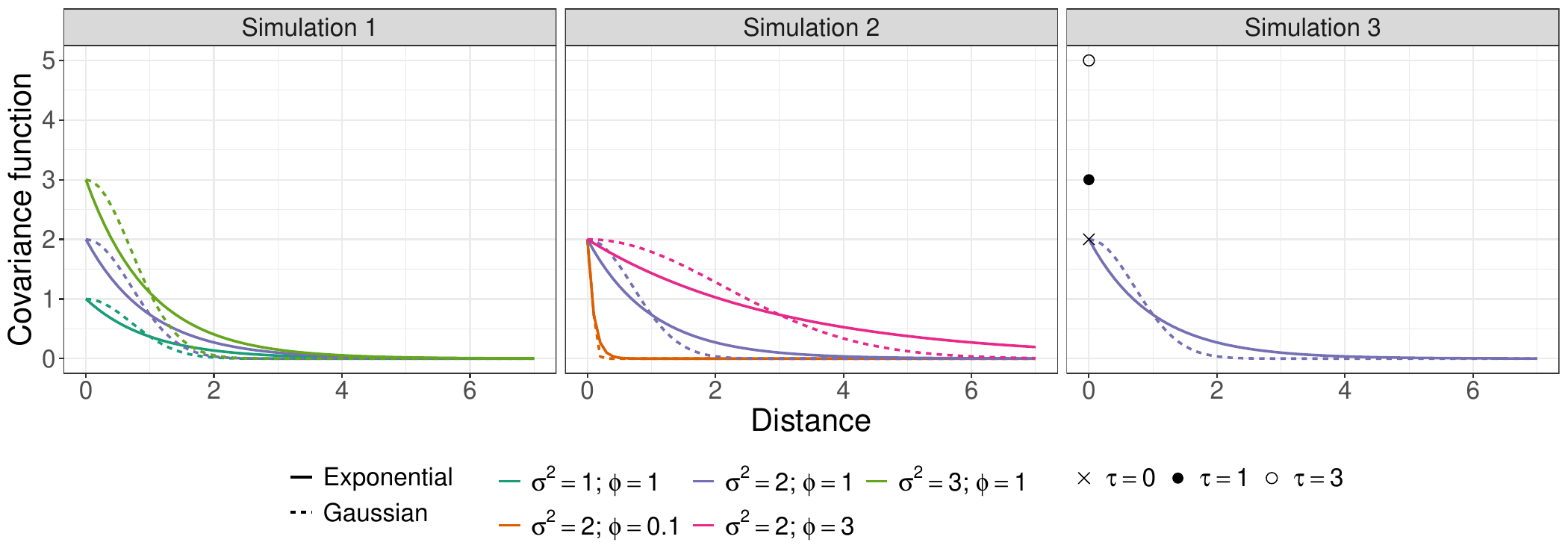}
    \caption{Covariance functions used in the three simulation experiments. Line styles represent the two covariance models, while colors indicate the different parameter settings. The left panel examines the impact of varying $\sigma^2_1$ and $\sigma^2_2$ while keeping $\phi_1 = \phi_2 = 1$. The central panel displays the effect of varying $\phi_1$ and $\phi_2$ while maintaining $\sigma^2_1 = \sigma^2_2 = 2$.  The right panel displays the effect of varying the nugget effect parameter when $\sigma^2_1 = \sigma^2_2 = 2$ and $\phi_1 = \phi_2 = 1$, resulting in different discontinuity points in zero.}
    \label{fig:simulation_kernels}
\end{figure}

In the first simulation experiment, we investigate the effect of changes in the variance parameters while keeping the scale parameters constant. On the contrary, we vary the scale parameters in the second experiment while holding the variances constant. Finally, the third scenario introduces three levels of nugget effects into the covariance functions. Additionally, we explore how nearly equally sized clusters can be distinguished from unbalanced clusters. Specifically, we consider the cases where $N_1 = N_2 = 100$ and $N_1 = 150$, $N_2 = 50$. For each combination of cluster sizes and covariance parameters, we generate ten datasets as follows. First, we randomly sample 500 spots within the square $(0, 10) \times (0, 10)$. Then, for each spot, we simulate 200 observations.

We estimate our SR-GMM by running the estimation algorithm ten times with random initializations on each dataset, retaining the parameter estimates that yielded the highest posterior density value. We compare the performance of our SR-GMM with two other clustering models: k-means,  randomly initialized multiple times, and the parsimonious GMM, initialized using hierarchical clustering and selected based on the BIC criterion, as implemented in the \textsf{R} package \texttt{mclust}  \citep{Scrucca_etal.2016}. The purpose of these simulation experiments is to evaluate how effectively the models can recover the true clustering structure of the data under different covariance structures. To achieve this, we measure the discrepancy between the true and estimated partitions using the Rand index \citep{Rand.1971}. A Rand index of 1 indicates perfect agreement between the true and the estimated partitions, whereas the lower its value, the greater the dissimilarity. 

All analyses are conducted on a Linux server running Ubuntu 22.04.5 LTS (Jammy Jellyfish). The system is equipped with an Intel Xeon Gold 6226R CPU at 2.90 GHz (32 cores), 92 GB of RAM, and 8 GB of swap space.

\begin{figure}[t]
    \centering
    \includegraphics[width=0.45\linewidth]{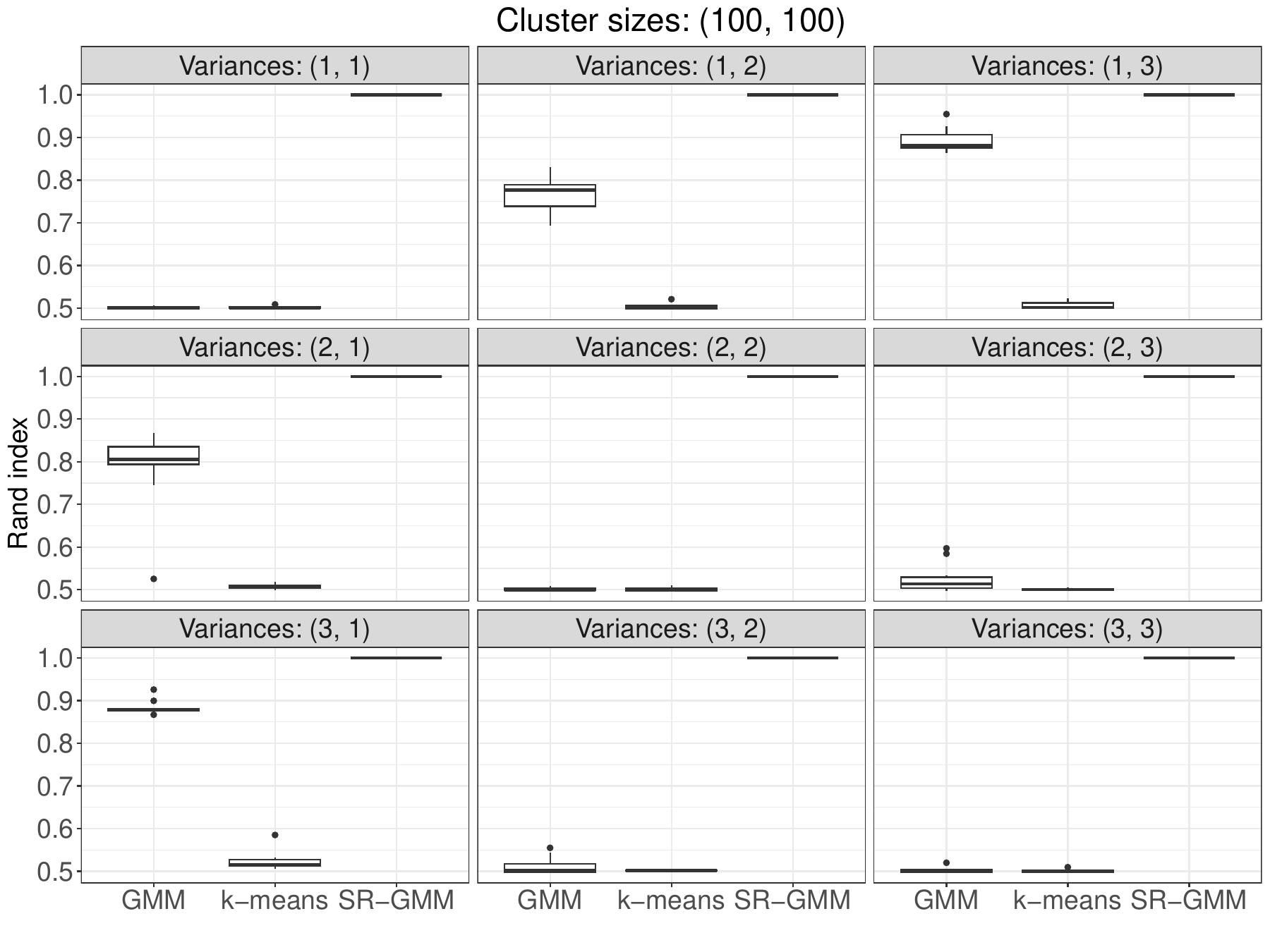}
    \includegraphics[width=0.45\linewidth]{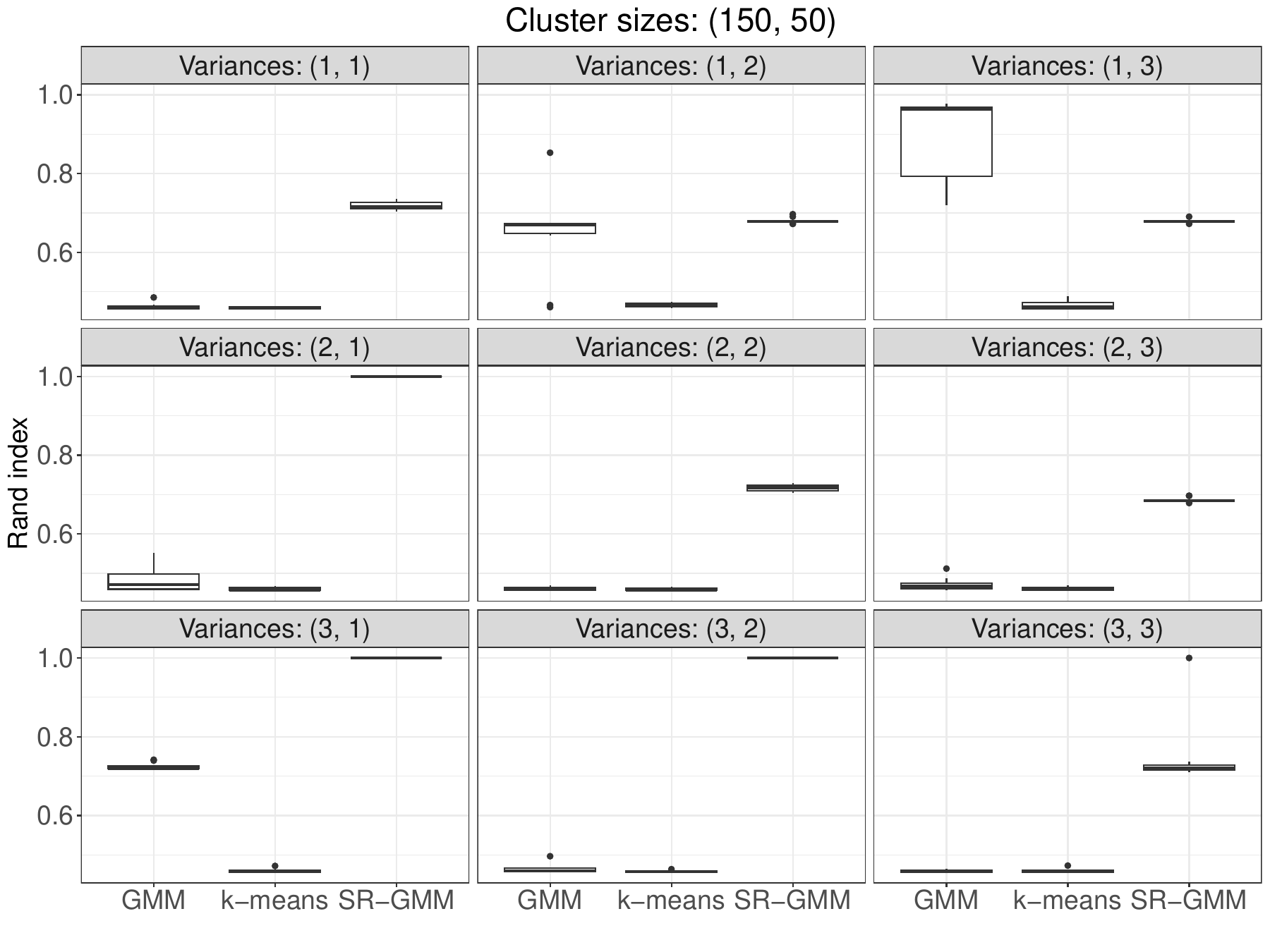}
    \includegraphics[width=0.45\linewidth]{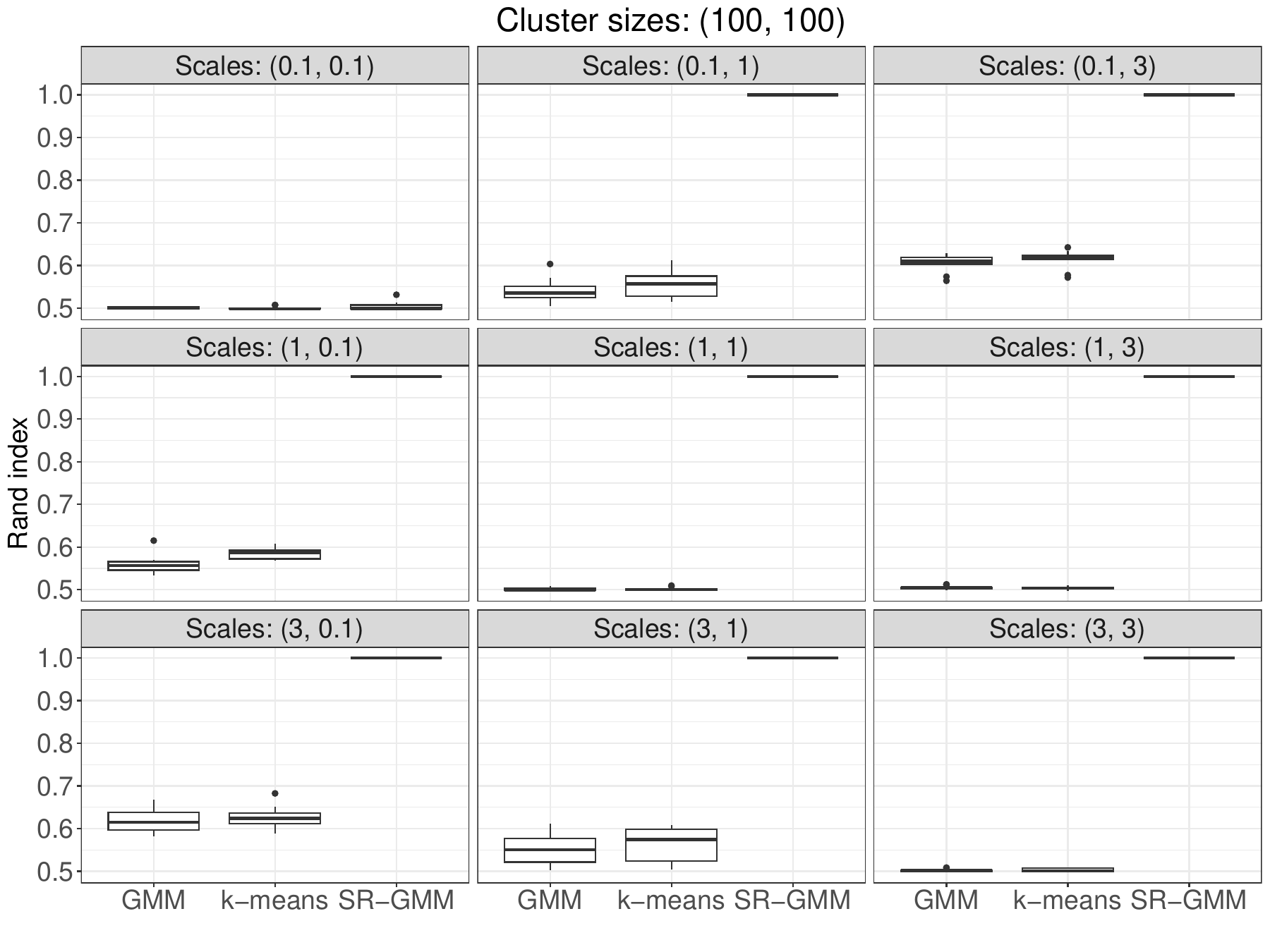}
    \includegraphics[width=0.45\linewidth]{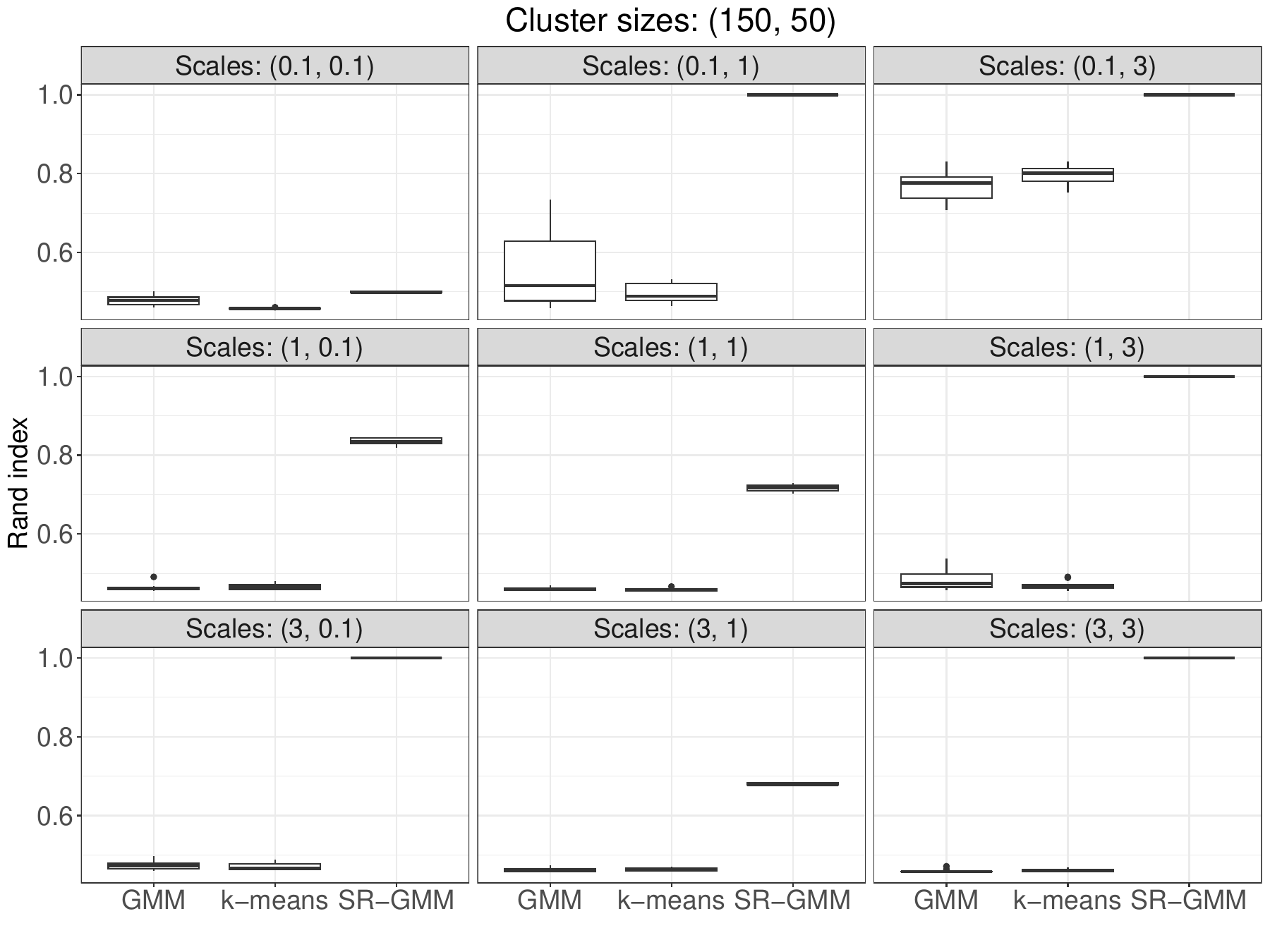}
    \includegraphics[width=0.45\linewidth]{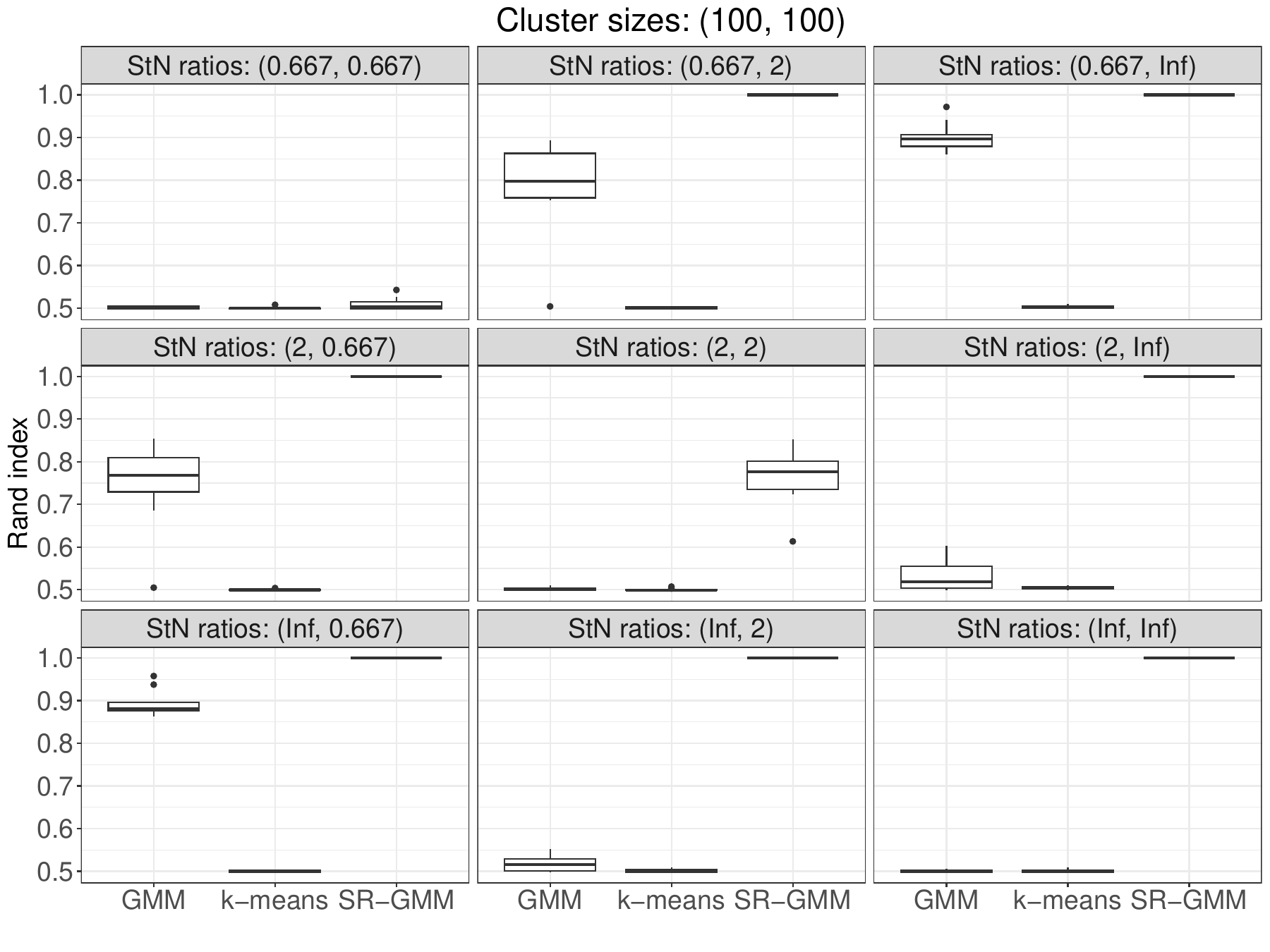}
    \includegraphics[width=0.45\linewidth]{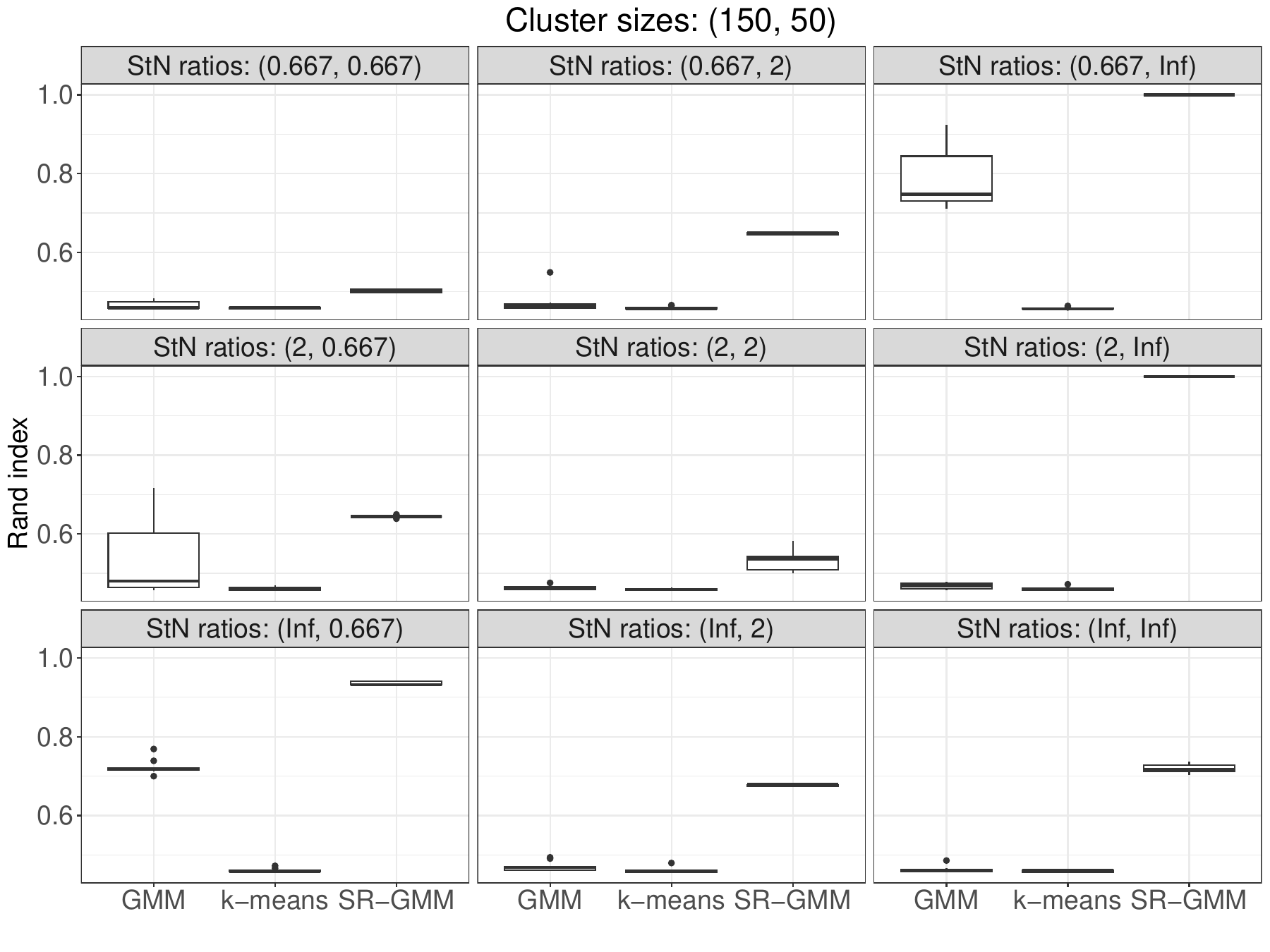}
    \caption{Results from the simulation experiments in the balanced framework (left) and in the unbalanced framework (right). The top row shows the results varying the clusters' variance parameters, the central row shows the results varying the clusters' scales, and the bottom row shows the results adding nugget effects. Each scenario is evaluated using 10 simulated datasets. Recall that a Rand index of 1 indicates perfect agreement between the true and the estimated partitions, whereas lower values correspond to greater dissimilarity.}
    \label{fig:simulation_results}
\end{figure}

\subsection{Impact of the variance parameter}
\label{subsec:simulation_variance}

First, we study the impact of different values of the variances $\sigma^2_1$ and $\sigma^2_2$. A representation of the exponential and Gaussian kernels for $\sigma^2_1, \sigma^2_2 \in \{1, 2, 3\}$ is provided in the left panel of Figure \ref{fig:simulation_kernels}. Simulation results are displayed in the top row of Figure \ref{fig:simulation_results}.

When the cluster sizes are balanced (top-left panel), our SR-GMM consistently retrieves the correct partition of the data. In contrast, the competing models fail to recover the clustering structure when $\sigma^2_1 = \sigma^2_2$, with their performance improving as the difference between the variances increases. These results suggest that both kernels are equally distinguishable, as all three models yield similar outcomes when either the exponential variance $\sigma^2_1$ or the Gaussian variance $\sigma^2_2$ is increased. This observation is supported by the near symmetry of the $3 \times 3$ matrix of panels in the top-left plot. The boxplots in panel (1,2) closely resemble those in panel (2,1), as do other symmetrically placed panels. This symmetry highlights the comparable behavior of the three estimated models, regardless of whether the data are generated with an exponential or with a Gaussian kernel.

In the case of unbalanced clusters (top-right panel), SR-GMM performs well overall, with a minimum Rand index of approximately 0.7 and a peak of 1. We observe that the model performs particularly well when $\sigma^2_1$, the parameter associated with the larger cluster, is larger than $\sigma^2_2$. Overall, our model outperforms the competing models, except in the case where $\sigma^2_1 = 1$ and $\sigma^2_2 = 3$, in which the smaller cluster exhibits a much larger variance. In this scenario, SR-GMM tends to estimate two balanced clusters, misclassifying some observations of Cluster 1. The parsimonious mixture model provides good but highly variable results in this context. In all cases, k-means does not appear to be a suitable solution.

The computational cost of estimating SR-GMM under this simulation setup -- measured by the average runtime per iteration and the number of iterations needed for convergence -- is shown in the top row of Supplementary Figure 1. Overall, the computational burden per iteration is lower when the two clusters differ in size (150 and 50 observations) compared to when they are equal in size (100 and 100 observations). However, when the smaller cluster (50 observations) has a variance equal to or greater than that of the larger cluster (150 observations), the algorithm generally requires more iterations to converge. These cases are represented by the boxplots labeled ``(150, 50)" appearing in the upper-triangular part of the top-right panel in Supplementary Figure 1. This effect is not observed when the clusters are of equal size.

\subsection{Impact of the scale parameter}
\label{subsec:simulation_scale}

Next, we examine the impact of varying the scale parameters $\phi_1$ and $\phi_2$ while keeping $\sigma^2_1 = \sigma^2_2 = 2$. We consider three scale values: 0.1, 1, and 3. The first value is chosen specifically to represent the case of very short spatial dependence. A visualization of the kernels generated under these configurations is shown in the central panel of Figure~\ref{fig:simulation_kernels}, and the simulation results are displayed in the central row of Figure \ref{fig:simulation_results}.

Both when the cluster sizes are balanced (central-left plot) and unbalanced (central-right plot), our SR-GMM consistently outperforms the competing models. In the balanced case, SR-GMM struggles to recover the clustering structure only when both clusters are characterized by scale values that are nearly zero. This is because, when the scale values are close to zero, the true covariance matrices of the clusters are approximately $\bSigma^{(1)}=\sigma^2_1\mathbf{I}_n$ and $\bSigma^{(2)}=\sigma^2_2\mathbf{I}_n$. In addition, since $\sigma^2_1 = \sigma^2_2$, the observations are effectively generated from the same distribution. However, as the scale of at least one cluster increases, then our model fully recovers the clustering structure. For GMM and k-means, the quality of clustering improves slightly as the difference in scale values between clusters increases; however, these models never achieve a Rand index greater than 0.7. Moreover, when the scales are equal, they fail to recover the original data partition entirely, as evident from the panels along the main diagonal of the central-left plot in Figure \ref{fig:simulation_results}.

In the unbalanced case (central-right plot), SR-GMM fails to perfectly recover the clustering structure in scenarios where the scale of the larger cluster, $\phi_1$, exceeds that of the smaller cluster, $\phi_2$. In contrast, when $\phi_2 > \phi_1$, the performance of all three models improves. These latter scenarios are particularly relevant for evaluating the performance of SR-GMM in contexts similar to real spatial transcriptomic experiments. In fact, spatial transcriptomic experiments are often characterized by a small subset of genes that exhibit large spatial variation, while the majority display only short-range spatial variation. The good results achieved by SR-GMM in these cases highlight its suitability for clustering gene expression patterns measured across thousands of spatial locations.

The computational cost of estimating SR-GMM under this simulation setup is shown in the central row of Supplementary Figure 1. Overall, the computational burden per iteration increases with the value of the scale parameter, and it is generally higher in scenarios with equally sized clusters (100 and 100 observations) compared to scenarios with unbalanced clusters (150 and 50 observations). The number of iterations required to reach convergence is generally small and similar across both balanced and unbalanced cluster configurations. The only exception is the case where both clusters have a scale of 0.1, which corresponds, as discussed above, to a scenario with virtually no spatial effect. In this setting, although the computational cost per iteration remains small, the algorithm requires a substantially larger number of iterations to converge.

\subsection{Addition of nugget effects}
\label{subsec:simulation_nugget}

Lastly, we investigate the impact of nugget effects in the determination of clusters. These quantities, denoted as $\tau_1$ and $\tau_2$, are added to the spatial kernel $\psi_1$ in Formula \eqref{formula:simulation_kernels} as $\tau_1 \mathds{1}(d = 0)$, and to the kernel $\psi_2$ as $\tau_2 \mathds{1}(d = 0)$. We consider the case $\sigma^2_1 = \sigma^2_2 = 2$ and $\phi_1 = \phi_2 = 1$, while $\tau_1, \tau_2 \in \{0, 1, 3\}$. 
The resulting covariance functions present a discontinuity point in 0, which equals $\sigma^2_1 + \tau_1$ for the exponential kernel and $\sigma^2_2 + \tau_2$ for the Gaussian kernel. We have depicted them in the right panel of Figure~\ref{fig:simulation_kernels}.
Since the nugget values should not be interpreted in absolute terms but in relation to the $\sigma^2$ values, we report the simulation results in terms of the signal-to-noise ratio, defined as $\sigma^2/\tau$, that is the ratio between the spatial variability and the nugget effect. Therefore, the settings considered in our simulations result in signal-to-noise ratios of 0.67, 2, and infinity. The latter corresponds to the case of no nugget effect. The simulation results are displayed in the bottom row of Figure \ref{fig:simulation_results}.

In the case of balanced clusters (bottom-left plot), SR-GMM perfectly recovers the clustering structure of the data for all combinations of signal-to-noise ratios, except for the cases $(0.667,0.667)$ and $(2,2)$. The first case represents a scenario where the nugget is more prevalent than the spatial variance in both clusters, significantly reducing the spatial covariance \citep[Section 8.2]{Waller_Gotway.2004}. As expected, the results in this case are consistent with those observed in Section \ref{subsec:simulation_scale} for the setting $\sigma^2_1 = \sigma^2_2 = 2$ and $\phi_1 = \phi_2 = 0.1$, which is the configuration with small scale values and no nugget effects. In the second case, the nugget is half the spatial covariance, resulting in a median Rand index greater than 0.7, with a maximum observed over the ten simulations exceeding 0.8.

The results are similar in the case of unbalanced clusters (bottom-right plot). Here, SR-GMM perfectly retrieves the clusters when the nugget only affects the exponential cluster, which is the most predominant. When the nugget is present in the less predominant cluster, SR-GMM still outperforms the competing models, but it does not fully recover the clustering structure. Finally, we recall that when both signal-to-noise ratios are infinite, the nugget effects are null, and the results on balanced and unbalanced clusters correspond to those shown in the central panels of the top-row figures.

The computational cost of estimating SR-GMM under this simulation setup is shown in the bottom row of Supplementary Figure 1. Overall, the algorithm requires more iterations to converge when the signal-to-noise ratios are equal between the two clusters, except in the case where both are infinite—corresponding to the absence of a nugget effect. Another factor that increases the number of iterations is when the clusters are unbalanced and the smaller cluster has a non-zero nugget effect. These cases are illustrated in the first two columns of the bottom-right panel in Supplementary Figure 1. The computational burden per iteration remains low and fairly stable across all settings.

We would like to highlight one particular aspect that emerged from this simulation experiment. When we applied SR-GMM, we ran the estimation algorithm 10 times, and we retained the results corresponding to the highest values of the posterior density. These are the results that appear in the bottom row of Figure \ref{fig:simulation_results}. However, when the model is misspecified due to the presence of a nugget effect, this approach may not be optimal. In some cases, across the 10 runs, the clustering that best aligns with the true data structure does not come from the model achieving the highest posterior density value. This result highlights the need for future developments to explore more accurate criteria for model selection, particularly in the presence of model misspecification due to non-negligible nugget effects in the data.

\section{Application to prostate cancer data}
\label{sec:application}

\begin{figure}[t]
    \centering
    \includegraphics[width=1\linewidth]{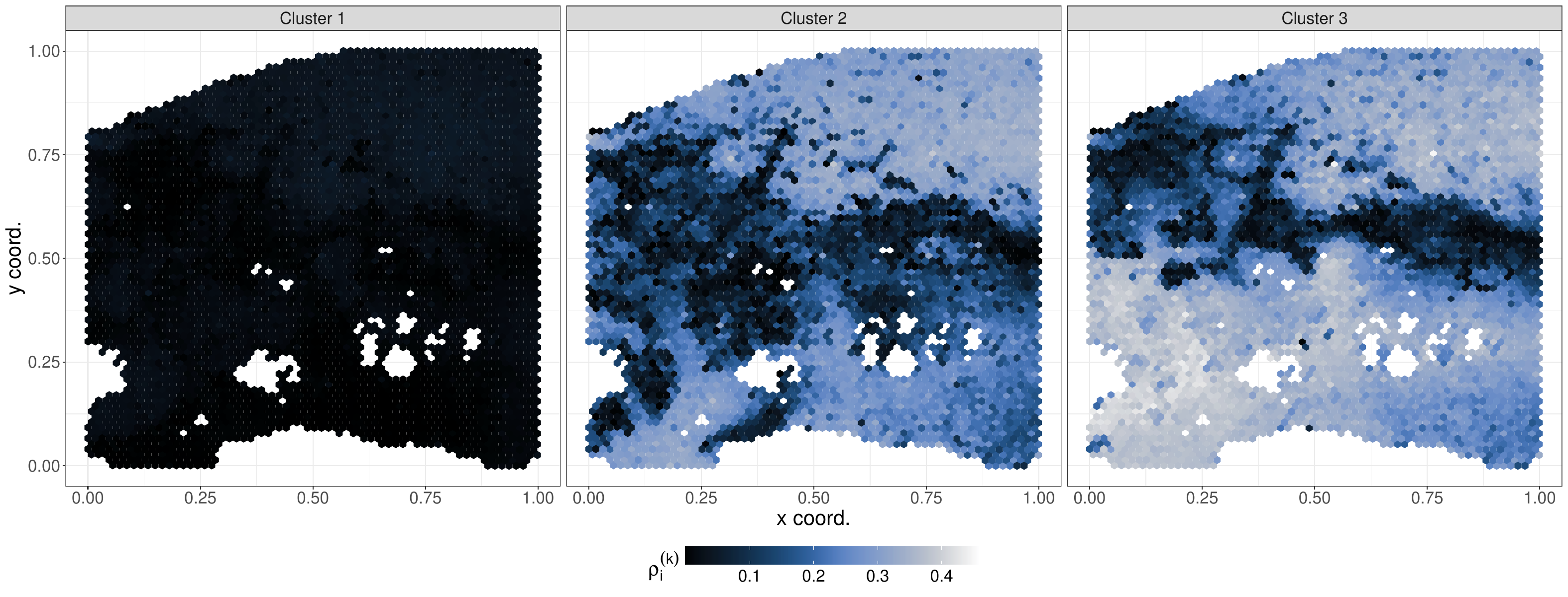}
    \caption{Results on the prostate cancer tissue sample. A generic spot $i$ in the $k$-th panel, with $k = 1,2,3$, is colored according to $\rho\pk_i$, as defined in Formula \eqref{formula:median_absolute_correlations}.}
    \label{fig:application_correlations}
\end{figure}

In this section, we apply SR-GMM to the real prostate cancer sample introduced in Section \ref{subsec:data_structure}. The dataset comprises the gene expressions of 500 genes across more than 4000 spots. We aim to identify clusters of genes that exhibit similar spatial correlation structures. 
We recall that genes are selected based on the deviance criterion proposed by \cite{Townes_etal.2019}. This method fits a multinomial model to each gene individually and computes the deviance, with higher deviance values indicating genes whose expression varies significantly across the tissue. Since deviance residuals are not well-defined for the multinomial model, \cite{Townes_etal.2019} proposed approximating the multinomial distribution with a binomial one. Accordingly, the top 500 highly variable genes are selected using this criterion, while the remaining genes are excluded from the analysis. For each selected gene $\ell$, we define $\mathbf{y}_{\ell,.}$ as the vector of deviance residuals obtained by fitting the multinomial model to the vector of counts of the $\ell$-th gene, thereby allowing us to work with continuous data. A representation of each vector $\mathbf{y}_{\ell,.}$ is given in Supplementary Figure 2. Notice that each $\mathbf{y}_{\ell,.}$ is not zero-centered. 
This entire pre-processing procedure is carried out using the \textsf{R} package \texttt{scry} \citep{Street_etal.2023}.

We conduct preliminary experiments with values of $K$ ranging from 3 to 5. For the sake of conciseness, we present the results obtained with $K=3$, which are also the most interpretable from a biological perspective.  
As already done in the simulation experiments presented in Section \ref{sec:simulations}, we estimate our SR-GMM by running the estimation algorithm ten times with random initializations, retaining the parameter estimates that yield the highest posterior density value. Details on the computational costs of the estimation process are given in Supplementary Figure 3. To present the results, we calculate the correlation matrices for each cluster through the following steps. First, we derive the maximum a posteriori estimates of the precision matrices, $(\bSigma\pk)^{-1}$, for $k = 1, 2, 3$, using the posterior distributions of $\mathbf{U}\pk$ and $\mathbf{D}\pk$. Then, we invert the precision matrices to obtain the covariance matrices $\bSigma\pk$. Finally, we use $\bSigma\pk$ to compute the correlation matrices $\mathbf{R}\pk$. 
For each spot $i$ within cluster $k$, we compute the following measure:  
\begin{equation}
    \label{formula:median_absolute_correlations}
\rho\pk_i = \mathrm{med}(|\mathbf{R}\pk_{i,-i}|),
\end{equation}
where $\mathrm{med}(\mathbf{a})$ denotes the median of the elements in the vector $\mathbf{a}$, and $|\mathbf{R}\pk_{i,-i}|$ represents the vector of absolute correlations between spot $i$ and all other spots in the tissue within cluster $k$. Thus, $\rho\pk_i$ quantifies the median strength of association between the expression of the genes in the $k$-th cluster at spot $i$ and that at all the other spots. This measure, computed among the genes in the same cluster, summarizes the strength of the spatial correlations that is present between a spot and all the remaining ones.

In Figure~\ref{fig:application_correlations}, we represent the spatial distribution of $\rho\pk_i$ across the three clusters. The first cluster appears spatially inactive, indicating that it lacks a spatial structure. Identifying a cluster characterized by the absence of spatial activity serves as a heuristic stopping point, indicating that no additional clusters need to be added to the model.
In the second cluster, the areas identified as tumor and stroma exhibit higher spatial connectivity. At the same time, a region on the left side of the image shows a lower level of spatial connectivity. 
Lastly, in the third cluster, we observe a high level of spatial connectivity across a large portion of the tissue, except for a narrow region that separates the tumor in the top-right from the stroma in the lower part of the tissue. 

To further interpret the results, we conduct a gene set enrichment analysis \citep{khatri2012ten}. Gene sets are groups of genes associated with specific biological processes, also referred to as signatures. Identifying a cluster of genes enriched in those related to a specific signature provides valuable insights into the biological processes underlying the tissue under study.
We perform the enrichment analysis separately for each cluster $k$, comparing the number of genes associated with a specific signature to the total of 500 genes in the dataset associated with the same signature. A Fisher exact test is used to identify the signatures in which the genes of cluster $k$ are significantly enriched; p-values are adjusted for the false discovery rate \citep{Benjamini_Hochberg.1995}. 
For this analysis, we use the gene sets included in the \textit{hallmark gene sets} available on the \textit{Gene Set Enrichment Analysis} website\footnote{\url{https://www.gsea-msigdb.org/gsea/msigdb}}. These data are also accessible through the \textsf{R} package \texttt{msigdbr} \citep{Dolgalev.2022}.

Cluster 1 is found to be enriched in genes associated with the \textit{hallmark TNF-$\alpha$ signaling via NF-$\kappa$B} signature (adjusted p-value = 0.01). This result suggests that genes in this cluster are linked to the biological processes through which the pro-inflammatory cytokine TNF-$\alpha$ activates the NF-$\kappa$B pathway. The activation of this pathway is known to drive critical aspects of tumor progression, including the proliferation of tumor cells, inhibition of apoptosis, enhancement of angiogenesis, and facilitation of invasion and metastasis \citep{Tang_etal.2017}.

Cluster 2 is found to be enriched in genes associated with epithelial-mesenchymal transition (adjusted p-value = 0.035), a developmental process observed in cancer progression and metastasis \citep{gavert2008epithelial}. This state characterizes the connective cells that are associated with the tumor here and surround the tumor mass, one of the main functions of which is to modify the tumor microenvironment by reshaping the extracellular matrix and thus promoting tumor progression \citep{Belhabib_etal.2021}.

Lastly, Cluster 3 is not significantly enriched in genes associated with specific signatures. However, the \textit{hallmark androgen response} signature has the smallest adjusted p-value (0.113). This gene set represents the cellular response to androgens, critical for regulating prostate cell growth and function. In the context of prostate cancer, androgen response genes are of particular interest because androgen deprivation therapy, which reduces androgen levels or blocks androgen receptor activity, is a standard treatment to limit tumor progression \citep{Kregel_etal.2020}. Therefore, near-significant enrichment of genes from this signature suggests a potential connection between Cluster 3 and androgen-mediated processes.

The fact that the three clusters are specifically enriched for different biological processes critical for prostate cancer suggests that our clustering strategy highlights important gene groups whose spatial distribution differs. These results may serve as a basis for further investigation into the role of these processes in prostate cancer biology.

\section{Discussion}
\label{sec:discussion}
In this paper, we have extended the nonparametric, non-stationary model introduced by~\citet{kidd_katzfuss.2022} into a model-based clustering framework. Specifically, we have proposed estimating the mean level and the covariance structure of the data using clustered observations as repeated measurements. To enable efficient inference and obtain point estimates from the posterior, we integrated the marginalized estimation strategy of~\citet{kidd_katzfuss.2022} within a Stochastic EM algorithm. While the proposed model has effectively identified clusters of genes with similar covariance structures, this work lays the foundation for several promising extensions.
First, exploring alternative approximate inference techniques within a fully Bayesian framework, such as Variational Bayes~\citep{Blei2017}, would be worthwhile. Second, a notable limitation of the current approach is the absence of an automated procedure for determining the number of clusters, $K$. In our application, we constrained the number of fitted components to a small value to facilitate interpretability, guided by prior knowledge of the dataset. Developing a method that dynamically eliminates redundant clusters, independent of the dataset at hand, represents a compelling avenue for future research. In this regard, Bayesian nonparametric or sparse finite mixture models offer a promising direction. For instance, Dirichlet Process mixtures and their extensions, estimated via Variational Bayes~\citep{Blei2006,Dangelo2023}, could provide a flexible solution. Third, the model can be extended to more effectively account for the presence of nugget effects. This aspect significantly complicates inference, as previously discussed by~\cite{kidd_katzfuss.2022}. While our simulation experiments demonstrated that SR-GMM achieves good results also in the presence of various levels of nugget effects, we believe that explicitly incorporating this additional source of variability into the model could yield substantial improvements in both model fitting and clustering accuracy. 
Finally, it would be intriguing to adapt the proposed modeling framework for applications such as image segmentation, drawing inspiration from~\cite{Sottosanti_etal.2025}, or extending it to handle partially or separately exchangeable settings, as described in~\cite{Denti2021}, \cite{Lin2021}, and \cite{Denti2025}.

\subsubsection*{Data availability} 
The data used in the application are openly available at the website of \href{https://www.10xgenomics.com/datasets/human-prostate-cancer-adenocarcinoma-with-invasive-carcinoma-ffpe-1-standard-1-3-0}{10X Genomics}.

\subsubsection*{Supplementary Material} 
The Supplementary Material reports additional results on the computational cost of SR-GMM and provides an overview of the gene expression profiles in the prostate cancer tissue sample analyzed in the manuscript. All the analyses presented in this article can be  reproduced using the material in the repository \url{https://github.com/andreasottosanti/SR-GMM_JoC.git}.

\subsubsection*{Acknowledgments}
Davide Risso was supported by EU funding within the MUR PNRR ``National Center for HPC, big data and quantum computing'' (Project no. CN00000013 CN1). The views and opinions expressed are only those of the authors and do not necessarily reflect those of the European Union or the European Commission. Neither the European Union nor the European Commission can be held responsible for them.
Andrea Sottosanti and Francesco Denti were supported by two distinct SID research projects awarded by the Department of Statistical Sciences of the University of Padova. The projects name are ``Flexible and interpretable mixture models with spatially dependent weights'' and ``CAPAREXTRA: Common Atoms Priors and Algorithms performing Reliable and
Efficient inference analyzing nested data with compleX TRAits'', respectively. Andrea Sottosanti was also supported by funding from the Italian Ministry of University and Research under the Call for Proposals related to the scrolling of the final rankings of the PRIN 2022 call (Project title “Latent variable models and dimensionality reduction methods for complex data”, Project No. 20224CRB9E, CUP C53C24000730006, PI Prof. Paolo Giordani, Grant Assignment Decree No. 1401 adopted on 18.9.2024 by MUR).

\bibliography{sn-bibliography}

\end{document}